
\pretolerance 50
\hsize=3.2in 
\vsize=9.1in 
\output{\plainoutput
  \ifnum\pageno=2 \global\vsize=6.65in
  \else \global\vsize=9.1in
  \fi}
\tolerance=1000
\def\newsection#1{\bigbreak\bigskip\bigskip\noindent{\twelvebf\spaceskip=.4em
    \xspaceskip=.6em #1}\par
    \nobreak\medskip\vskip2pt\noindent\ignorespaces}
\font\tenrm=ptmr at 10.5pt
\font\sevenrm=ptmr at 7pt
\font\fiverm=ptmr at 5pt
\textfont0=\tenrm \scriptfont0=\sevenrm \scriptscriptfont0=\fiverm
\font\elevenbf=ptmb at 11pt
\font\twelvebf=ptmb at 12pt
\font\twelvett=cmtt12

\rm
\hyphenchar\tentt=-1

\font\em=ptmri at 10.5pt

\font\ninerm=ptmr at 9pt

\let\mc=\ninerm
 \let\sc=\eightrm
\font\ninebf=ptmb at 8.9pt

\font\logosl=logosl10
\font\logo=logo10
\font\sltt=cmsltt10
\font\title=ptmb at 16.5pt

\def\CTWILL/{{\tt CTWILL}}
\def\CWEB/{{\tt CWEB}}
\def\slMF{{\logosl METAFONT}}
\def\MF{{\logo METAFONT}}
\newbox\PPbox 
\setbox\PPbox=\hbox{\kern.5pt\raise1pt\hbox{\sevenrm+\kern-1pt+}\kern.5pt}
\def\PP{\copy\PPbox}
\newbox\MMbox \setbox\MMbox=\hbox{\kern.5pt\raise.5pt\hbox{\ninerm-%
 \kern0pt-}\kern.5pt}
\def\MM{\copy\MMbox}
\newbox\MGbox 
\setbox\MGbox=\hbox{\kern-2pt\lower3pt\hbox{\teni\char'176}\kern1pt}
\def\MG{\copy\MGbox}
\def\CPLUSPLUS/{{\mc C\PP\spacefactor1000}}
\def\CEE/{{\mc C\spacefactor1000}}
\def\tt{\chardef\_=`\_\chardef\\=`\\\chardef\{=`\{\chardef\}=`\}\tentt}
\def\\#1{\leavevmode\hbox{\it#1\/\kern.05em}} 
\def\,{\relax\ifmmode\mskip\thinmuskip\else\thinspace\fi}
\let\*=\\
\def\TeX{{\ifmmode\it\fi
   \leavevmode\hbox{T\kern-.1em\lower.424ex\hbox{E}\hskip-.075em X}}}

\def\HYP{1}
\def\WEB{2}
\def\TTP{3}
\def\MTP{4}
\def\LP{5}
\def\Cweb{6}
\def\SGB{7}
\def\Ctwill{8}
\def\SGBcode{9}
\def\Virg{10}
\def\Ween{11}

\vbox to 2.45in{\vfill 
\leftline{\title Mini-Indexes for Literate Programs}
\bigskip
\vskip3pt
\leftline{\elevenbf Donald E. Knuth}
\smallskip
\leftline{\ninerm Computer Science Department, Stanford University,
Stanford, CA 94305-2140 USA}
\bigskip
\bigskip}

{\ninerm\baselineskip=11pt\noindent
{\ninebf Abstract.}\enspace This paper describes how to implement
a documentation technique
that helps readers to understand large programs or collections of programs,
by providing local indexes to all identifiers that are visible on
every two-page spread. A detailed example is given for a program that
finds all Hamiltonian circuits in an undirected graph.
\bigskip
\bigskip
\vskip8pt
\hrule
\smallskip
\vskip4pt
\let\tt=\ninett
\noindent {\ninebf Keywords:}
 Literate programming, {\tt WEB}, \CWEB/, \CPLUSPLUS/,
\CTWILL/, \TeX, indexes, hypertext, Hamiltonian circuits
\smallskip
\vskip4pt
\hrule
\vskip8pt
}
\advance\baselineskip 0pt plus .1pt

\newsection{1. Introduction}
Users of systems like {\tt WEB} [\WEB], which provide support for
structured documentation and literate programming~[\LP], automatically get
a printed index at the end of their programs, showing where each identifier
is defined and used. Such indexes can be extremely helpful, but they can also
be cumbersome, especially when the program is long. An extreme example is
provided by the listing of \TeX~[\TTP], where the index contains 32 pages of
detailed entries in small print.

Readers of [\TTP] can still find their way around the program quickly,
however, because
{\smallskip\narrower \noindent
\dots the right-hand pages of this book contain mini-indexes
that will make it unnecessary for you to look at the big index very often.
Every identifier that is used somewhere on a pair of facing pages is listed
in a footnote on the right-hand page, unless it is explicitly defined or
declared somewhere on the left-hand or right-hand page you are reading.
These footnote entries tell you whether the identifier is a procedure or a
macro or a boolean, etc. \ [\TTP]
\smallskip}\noindent
A similar idea is sometimes used in
editions of literary texts for foreign language students, where
mini-dictionaries
of unusual words appear on each page~[\Virg]; this saves the student
from spending a lot of time searching big dictionaries.

The idea of mini-indexes was first suggested to the author by Joe Weening,
who prepared a brief mockup of what he thought might be possible~[\Ween].
His proposal was immediately appealing, so the author decided to implement
it in a personal program called {\tt TWILL}---a name suggested by the fact
that it was a two-pass variant of the standard program called {\tt WEAVE}.
\ {\tt TWILL} was used in September, 1985, to produce [\TTP] and a companion
book~[\MTP].

The original {\tt WEB} system was a combination of \TeX~and Pascal. But
the author's favorite programming language nowadays
is \CWEB/~[\Cweb], which
combines \TeX\ with~\CEE/. (In fact, \CWEB/ version 3.0 is fully compatible
with \CPLUSPLUS/, although the author usually restricts himself to a
personal subset that might be called \CEE/\MM.) One of the advantages of
\CWEB/ is that it supports collections of small program modules and
libraries that can be combined in many ways. A single \CWEB/ source file
{\tt foo.w} can generate several output files in addition to the \CEE/
program {\tt foo.c}; for example, {\tt foo.w} might generate a header file
{\tt foo.h} for use by other modules that will be loaded with the object
code {\tt foo.o}, and it might generate a test program {\tt testfoo.c} that
helps verify portability.

\CWEB/ was used to create the Stanford GraphBase, a collection of about three
dozen public-domain programs useful for the study of combinatorial
algorithms~[\SGBcode]. These programs have recently been published in book
form, again with mini-indexes~[\SGB]. The mini-indexes in this case were
prepared with \CTWILL/~[\Ctwill], a two-pass variant of {\tt CWEAVE}.

The purpose of this paper is to explain the operations of {\tt TWILL} and of
its descendant, \CTWILL/. The concepts are easiest to understand when they
are related to a detailed example, so a complete \CWEB/ program has been
prepared for illustrative purposes. Section~2 of this paper explains the
example program; Sections 3 and~4 explain how \CTWILL/ and \TeX\ process
it; and Section~5 contains concluding comments.

\newsection{2. An example}
The \CWEB/ program for which sample mini-indexes have been prepared
especially for this paper is called {\sc HAM}. It enumerates all
Hamiltonian circuits of a graph, that is, all undirected cycles that
include each vertex exactly once. For example, the program can determine
that there are exactly 9862 knight's tours on a $6\times6$ chessboard,
ignoring symmetries of the board, in about 2.3 seconds on a SPARC\kern1pt
station~2.
Since {\sc HAM} may be interesting in its own right, it is presented in its
entirety as sort of a ``sideshow'' in the right-hand columns of the pages
of this article and on the final (left-hand) page.

Please take a quick look at {\sc HAM} now, before reading further.
The program appears in five columns, each of which will be called a
{\em spread\/} because it is analogous to the two-page spreads in [\TTP]
and~[\SGB]. This arrangement gives us five mini-indexes to look at
instead of just two, so it makes {\sc HAM} a decent example in spite
of its relatively small size. A shorter program wouldn't need much of an
index at all; a longer program would take too long to read.

{\sc HAM} is intended for use with the library of routines that comes with
the Stanford GraphBase, so \S1 of the program tells the \CEE/ preprocessor
to include header files {\tt gb\_graph.h} and {\tt gb\_save.h}. These
header files define the external functions and data types needed from the
GraphBase library.

A brief introduction to GraphBase data structures will suffice for the
interested reader to understand the full details of {\sc HAM}. A graph is
represented by combining three kinds of {\bf struct} records
called {\bf Graph}, {\bf Vertex}, and {\bf Arc}. If $v$ points to a
{\bf Vertex} record, $v\MG\\{name}$ is a string that names the vertex
represented by~$v$, and $v\MG\\{arcs}$ points to the representation of the
first arc emanating from that vertex.\footnote{$^1$}{`$v\MG\\{name}$'
 is actually typed `{\tt v->name}' in a
\CEE/ or \CWEB/ program; typographic sugar makes the program easier to read
in print.} If $a$ points to an {\bf Arc} record
that represents an arc from some vertex~$v$ to another vertex~$u$, then
$a\MG\\{tip}$ points to the {\bf Vertex} record that represents~$u$; also
$a\MG\\{next}$ points to the representation of the next arc from~$v$, or
$a\MG\\{next}=\Lambda$ (i.e., {\tt NULL}) if $a$ is the last arc from~$v$.
Thus the following loop will print the names of all vertices adjacent to~$v$:
$$\vbox{\halign{#\hfil\cr
{\bf for} $(a=v\MG\\{arcs};\ a;\ a=a\MG\\{next})$\cr
\quad\\{printf}({\tt"\%s\\n"},\,$a\MG\\{tip}\MG\\{name}$);\cr}}$$
An undirected edge between vertices $u$ and~$v$ is represented by two arcs,
one from $u$ to~$v$ and one from $v$ to~$u$. Finally, if $g$ points to a
{\bf Graph} record, then $g\MG n$ is the number of vertices in the
associated graph, and the {\bf Vertex} records representing those vertices
are in locations $g\MG\\{vertices}+k$, for $0\le k<g\MG n$.

A {\bf Vertex} record also contains ``utility fields'' that can be
exploited in different ways by different algorithms. The actual \CEE/
declarations of these fields, quoted from \S8 and \S9 of the program {\sc
GB\_\,GRAPH}~[\SGB], are as follows:
$$\vbox{\halign{#\hfill\cr
{\bf typedef\/ union} $\{$\cr
\quad {\bf struct} \\{vertex\_struct} $*V$;\cr
 \qquad$/*\,$ pointer to {\bf Vertex} $\,*/$\cr
\quad {\bf struct} \\{arc\_struct} $*A$;\cr
 \qquad$/*\,$ pointer to {\bf Arc} $\,*/$\cr
\quad {\bf struct} \\{graph\_struct} $*G$;\cr
 \qquad$/*\,$ pointer to {\bf Graph} $\,*/$\cr
\quad {\bf char} $*S$;\cr
 \qquad$/*\,$ pointer to string $\,*/$\cr
\quad {\bf long} $I$;\cr
 \qquad$/*\,$ integer $\,*/$\cr
$\}$ {\bf util}\kern1pt;\cr
\noalign{\smallskip}
{\bf typedef\/ struct} \\{vertex\_struct} $\{$\cr
\quad {\bf struct} \\{arc\_struct} $*\\{arcs}$;\cr
 \qquad$/*\,$ linked list of arcs out of this vertex $\,*/$\cr
\quad {\bf char} $*\\{name}$;\cr
 \qquad$/*\,$ string identifying this vertex symbolically $\,*/$\cr
\quad {\bf util} $u,v,w,x,y,z$;\cr
 \qquad$/*\,$ multipurpose fields $\,*/$\cr
$\}$ {\bf Vertex}\kern1pt;\cr}}$$

Program {\sc HAM} uses the first four utility fields in order to do its
word efficiently. Field~$u$, for example, is treated as a {\bf long}
integer representing the degree of the vertex. Notice the definition of
\\{deg} as a macro in~\S2; this makes it possible to refer to the degree
of~$v$ as $v\MG\\{deg}$ instead of the more cryptic `$v\MG u.I$' actually
seen by the \CEE/ compiler. Similar macros for utility fields $v$, $w$,
and~$x$ can be found in \S4 and~\S6.

\looseness=-1
The first mini-index of {\sc HAM}, which can be seen
below \S2 in the first column of the
program, gives cross-references to all identifiers that appear in \S1~or~\S2
but are not defined there. For example, \\{restore\_graph} is
mentioned in one of the comments of \S1; the mini-index tells us that it is
a function, that it returns a value of type {\bf Graph}~$*$, and that it is
defined in \S4 of another \CWEB/ program called {\sc GB\_\,SAVE}. The
mini-index also mentions that {\bf Vertex} and \\{arcs} are defined in \S9
of {\sc GB\_\,GRAPH} (from which we quoted the relevant definitions
above), and that fields
\\{next} and \\{tip} of {\bf Arc} records are defined in {\sc
GB\_\,GRAPH}~\S10, etc.

One subtlety of this first mini-index is the entry for~$u$, which tells us
that $u$ is a utility field defined in {\sc GB\_\,GRAPH}~\S9. The
identifier~$u$ actually appears twice in \S2, once in the definition of
\\{deg} and once as a variable of type {\bf Vertex}~$*$. The mini-index
refers only to the former, because the latter usage is defined in~\S2.
Mini-indexes don't mention identifiers defined within their own spread.

The second mini-index, below \S5 of {\sc HAM}, is similar to the first.
Notice that it contains two separate entries for~$v$, because the
identifier~$v$ is used in two senses---both as a utility field (in the
definition of \\{taken}) and as a variable (elsewhere). The \CEE/ compiler
will understand how to deal with constructions like `$v\MG v.I=0$', which
the \CEE/ preprocessor expands from `$v\MG\\{taken}=0$', but human readers
are spared such trouble.

Notice the entry for \\{deg} in this second mini-index: It uses an equals
sign instead of a colon, indicating that \\{deg} is a macro rather than a
variable. A~similar notation was used in the first mini-index for
cross-references to typedef'd identifiers like {\bf Vertex}\kern1pt.
 See also the
entry for \\{not\_taken} in the fourth mini-index: Here
`$\\{not\_taken}=\rm macro$~(\,)' indicates that \\{not\_taken} is a macro
with arguments.

\newsection{3. The operation of \twelvett \rlap{\kern.002in CTWILL}CTWILL}
It would be nice to report that the program \CTWILL/ produces the
mini-indexes for {\sc HAM} in a completely automatic fashion, just as {\tt
CWEAVE} automatically produces ordinary indexes. But that would be a lie.
The truth is that \CTWILL/ only does about 99\% of the work automatically;
the user has to help it with the hard parts.

Why is this so? Well, in the first place, \CTWILL/ isn't smart enough to
figure out that the `$u$' in the definition of \\{deg} in \S2 is not the
same as the `$u$' declared to be {\bf register Vertex}~$*$ in that same
section. Indeed, a high degree of artificial intelligence would be required
before \CTWILL/ could deduce that.

In the second place, \CTWILL/ has no idea what mini-index entry to make for
the identifier $k$ that appears in \S6. No variable $k$ is declared
anywhere! Indeed, users who write comments involving
expressions like `$f(x)$' might or might not be referring to identifiers
$f$ and/or~$x$ in their programs; they must tell \CTWILL/ when they are
making ``throwaway'' references that should not be indexed. {\tt CWEAVE}
doesn't have this problem because it indexes only the definitions, not the
uses, of single-letter identifiers.

In the third place, \CTWILL/ will not recognize automatically that the
\\{vert} parameter in the definition of \\{not\_taken}, \S4, has no
connection with the \\{vert} macro defined in~\S6.

A fourth complication, which does not arise in {\sc HAM} but does occur
in [\TTP] and [\SGB], is that sections of a {\tt WEB} or {\tt CWEB} program
can be used more than once. Therefore a single
identifier might actually refer to several different variables
simultaneously. (See, for example, \S652 in [\TTP].)

In general, when an identifier is defined or declared exactly once, and
used only in connection with its unique definition, \CTWILL/ will have no
problems with it. But when an identifier has more than one implicit or
explicit definition, \CTWILL/ can only guess which definition was meant.
Some identifiers---especially single-letter ones like $x$ and~$y$---are too
useful to be confined to a single significance throughout a large
collection of programs. Therefore \CTWILL/ was designed to let users
provide hints easily when choices need to be made.

The most important aspect of this design was to make \CTWILL/'s default
actions easily predictable. The more ``intelligence'' we try to build into
a system, the harder it is for us to control it. Therefore \CTWILL/ has
very simple rules for deciding what to put in mini-indexes.

Each identifier has a unique {\em current meaning}, which consists of three
parts: its type, and the program name and section number where it was
defined. At the beginning of a run, \CTWILL/ reads a number of
files that define the
initial current meanings. Then, whenever \CTWILL/
 sees a \CEE/ construction that
implies a change of meaning---a macro definition, a variable declaration, a
typedef, a function declaration, or the appearance of a label followed by a
colon---it assigns a new current meaning as specified by the semantics
of~\CEE/. For example, when \CTWILL/ sees `{\bf Graph} $*g$' in \S2 of {\sc
HAM}, it changes the current meaning of $g$ to `{\bf Graph}~$*$,~\S2'.
These changes occur in the order of the \CWEB/ source file, not in the
``tangled'' order that is actually presented to the \CEE/ compiler.
Therefore \CTWILL/ makes no attempt to nest definitions according to block
structure; everything it does is purely sequential. A variable declared in
\S5 and \S10 will be assumed to have the meaning of \S5 in \S6, \S7, \S8,
and~\S9.

Whenever \CTWILL/ changes the current meaning of a variable, it outputs a
record of that current meaning to an auxiliary file. For the \CWEB/ program
{\tt ham.w}, this auxiliary file is called {\tt ham.aux}. The first few
entries of {\tt ham.aux} are
$$\vbox{\halign{\tt#\hfil\cr
@\$deg \{ham\}2 =macro@>\cr
@\$argc \{ham\}2 \\\&\{int\}@>\cr
@\$argv \{ham\}2 \\\&\{char\} \$\{*\}[\\,]\$@>\cr
}}$$
and the last entry is
$$\hbox{\tt@\$d \{ham\}8 \\\&\{register\} \\\&\{int\}@>}$$
In general these entries have the form
$$\hbox{\tt@\$\*{ident} \{\*{name}\}\*{nn} \*{type}@>}$$
where \\{ident} is an identifier, \\{name} and \\{nn} are the program name
and section number where \\{ident} is defined, and \\{type} is a string of
\TeX\ commands to indicate its type. In place of
`{\tt\{\*{name}\}\*{nn}}' the entry might have the form {\tt"string"}
instead; then the program name and section number are replaced by the
string. (This mechanism leads, for example, to the appearance of
{\tt<stdio.h>} in {\sc HAM}'s mini-index entries for
\\{printf}.) Sometimes the \\{type} field says `{\tt\\zip}'. This situation
doesn't arise in {\sc HAM}, nor does it arise very often in [\SGB]; but it
occurs, for example, when a preprocessor macro name has been defined
externally as in a {\tt Makefile}, or when a type is very complicated, like
{\tt FILE} in {\tt<stdio.h>}. In such cases the mini-index will
simply say `{\tt FILE}, {\tt<stdio.h>}', with no colon or
equals sign.

The user can explicitly change the current meaning by specifying
{\tt@\$\*{ident}} {\tt\{\*{name}\}\*{nn}} {\tt\*{type}@>} anywhere in a
\CWEB/ program. This means that \CTWILL/'s default mechanism is easily
overridden.

When \CTWILL/ starts processing a program {\tt foo.w}, it looks first for a
file named {\tt foo.aux} that might have been produced on a previous run.
If {\tt foo.aux} is present, it is read in, and the {\tt@\$...@>} commands
of {\tt foo.aux} give current meanings to all identifiers defined in {\tt
foo.w}. Therefore \CTWILL/ is able to know the meaning of an identifier
before that identifier has been declared---assuming that \CTWILL/ has been
run successfully on {\tt foo.w} at least once before, and assuming that the
final definition of the identifier is the one intended at the beginning
of the program.

\CTWILL/ also looks for another auxiliary file called {\tt foo.bux}. This
one is
not overwritten on each run, so it can be modified by the user. The purpose
of {\tt foo.bux} is to give initial meanings to identifiers that are not
defined in {\tt foo.aux}. For example, {\tt ham.bux} is a file containing
the two lines
$$\vbox{\halign{\tt#\hfill\cr
@i gb\_graph.hux\cr
@i gb\_save.hux\cr}}$$
which tell \CTWILL/ to input the files {\tt gb\_graph.hux} and {\tt
gb\_save.hux}. The latter files contain definitions of identifiers
that appear in the header files {\tt gb\_graph.h} and {\tt gb\_save.h},
which {\sc HAM} includes in \S1. For example, one of the lines of {\tt
gb\_graph.hux} is
$$\hbox{\tt@\$Vertex \{GB\\\_GRAPH\}9 =\\\&\{struct\}@>}$$
This line appears also in {\tt gb\_graph.aux}; it was copied by hand, using
a text editor, into
{\tt gb\_graph.hux}, because {\bf Vertex} is one of the identifiers defined
in {\tt gb\_graph.h}.

\CTWILL/ also reads a file called {\tt system.bux}, if it is present; that
file contains global information that is always assumed to be in the
background as part of the
current environment. One of the lines in {\tt system.bux} is, for example,
$$\hbox{\tt@\$printf "<stdio.h>" \\\&\{int\} (\\,)@>}$$

After {\tt system.bux}, {\tt ham.aux}, and {\tt ham.bux} have been input,
\CTWILL/ will know initial current meanings of almost all identifiers that
appear in {\sc HAM}. The only exception is $k$, found in \S6; its current
meaning is {\tt\\uninitialized}, and if the user does not take corrective
action its mini-index entry will come out as
$$\hbox{$k$: ???, \S0.}$$

Notice that $d$ is declared in \S4 of {\sc HAM} and also in \S8. Both of
these declarations produce entries in {\tt ham.aux}. Since \CTWILL/ reads
{\tt ham.aux} before looking at the source file {\tt ham.w}, and since
{\tt ham.aux} is read sequentially, the current meaning of $d$ will refer
to \S8 at the beginning of {\tt ham.w}. This causes no problem, because $d$
is never used in {\sc HAM} except in the sections where it is declared,
hence it never appears in a mini-index.

When \CTWILL/ processes each section of a program, it makes a list of all
identifiers used in that section, except for reserved words. At the end of
the section, it mini-outputs (that is, it outputs to the mini-index) the
current meaning of each identifier on the list, unless that current meaning
refers to the current section of the program, or unless the user intervenes.

The user has two ways to change the mini-outputs, either by suppressing
the default entries or by inserting replacement entries. First, the explicit
command
$$\hbox{\tt@-\*{ident}@>}$$
tells \CTWILL/ not to produce the standard mini-output for \\{ident} in the
current section. Second, the user can specify one or more {\em temporary
meanings\/} for an identifier, all of which will be mini-output at the end
of the section. Temporary meanings do not affect an identifier's current
meaning. Whenever at least one temporary meaning is mini-output, the
current meaning will be suppressed just as if the {\tt@-...@>} command had
been given. Temporary meanings are specified by means of the operation
{\tt@\%}, which toggles a state switch affecting the {\tt@\$...@>} command:
At the beginning of a section, the switch is in ``permanent'' state, and
{\tt@\$...@>} will change an identifier's current meaning as described
earlier. Each occurrence of {\tt@\%} changes the state from ``permanent''
to ``temporary'' or back again; in ``temporary'' state the {\tt@\$...@>}
command specifies a temporary meaning that will be mini-output with no
effect on the identifier's permanent (current) meaning.

Examples of these conventions will be given momentarily, but first we
should note one further interaction between \CTWILL/'s {\tt@-} and {\tt@\$}
commands: If \CTWILL/ would normally assign a new current meaning to
\\{ident} because of the semantics of \CEE/, and if the command
{\tt@-\*{ident}@>} has already appeared in the current
section, \CTWILL/ will not
override the present meaning, but \CTWILL/ will output the present meaning
to the {\tt.aux} file. In particular, the user may have specified the present
meaning with {\tt@\$\*{ident}...@>}; this allows user control over what gets
into the {\tt.aux} file.

For example, here is a complete list of all commands inserted by the author
in order to correct or enhance \CTWILL/'s default mini-indexes for {\sc
HAM}:
\smallskip
\item{$\bullet$\ }At the beginning of \S2,
$$\tt\halign{#\hfil\cr
@-deg@>\cr
@\$deg \{ham\}2 =\\|u.\\|I@>\cr
@\%@\$u \{GB\\\_GRAPH\}9 \\\&\{util\}@>\cr
}$$
to make the definition of \\{deg} read `$u.I$' instead of just `macro' and
to make the mini-index refer to $u$ as a utility field.
\item{$\bullet$\ }At the beginning of \S4,
$$\tt\halign{#\hfil\cr
@-taken@> @-vert@>\cr
@\$taken \{ham\}4 =\\|v.\\|I@>\cr
@\%@\$v \{GB\\\_GRAPH\}9 \\\&\{util\}@>\cr
\ @\$v \{ham\}2 \\\&\{register\} \\\&\{Vertex\} \$*\$@>\cr
}$$
for similar reasons, and to suppress indexing of \\{vert}.
Here the mini-index gets two ``temporary'' meanings for~$v$, one of which
happens to coincide with its permanent meaning.
\item{$\bullet$\ }At the beginning of \S6,
$$\tt\halign{#\hfil\cr
@-k@> @-t@> @-vert@> @-ark@>\cr
@\$vert \{ham\}6 =\\|w.\\|V@>\cr
@\$ark \{ham\}6 =\\|x.\\|A@>\cr
@\%@\$w \{GB\\\_GRAPH\}9 \\\&\{util\}@>\cr
\ \ @\$x \{GB\\\_GRAPH\}9 \\\&\{util\}@>\cr
}$$
for similar reasons. That's all.
\smallskip
\noindent
These commands were not inserted into the program file {\tt ham.w}; they
were put into another file called {\tt ham.ch} and introduced via \CWEB/'s
``change file'' feature~[\Cweb]. Change files make it easy to modify the
effective contents of a master file without tampering with that file directly.

\newsection{4. Processing by \TeX}
\CTWILL/ writes a \TeX\ file that includes mini-output at the end of each
section. For example, the mini-output after \S10 of {\sc HAM} is
$$\tt\halign{#\hfil\cr
\\]\{GB\\\_GRAPH\}10 \\\\\{next\} \\\&\{Arc\} \$*\$\cr
\\[7 \\\\\{advance\} label\cr
\\[6 \\\\\{ark\} =\\|x.\\|A\cr
\\[2 \\|\{t\} \\\&\{register\} \\\&\{Vertex\} \$*\$\cr
\\[4 \\\\\{not\\\_taken\} =macro (\\,)\cr
\\]\{GB\\\_GRAPH\}10 \\\\\{tip\} \\\&\{Vertex\} \$*\$\cr
\\[2 \\|\{v\} \\\&\{register\} \\\&\{Vertex\} \$*\$\cr
\\[2 \\|\{a\} \\\&\{register\} \\\&\{Arc\} \$*\$\cr
}$$
Here {\tt\\[} introduces an internal reference to another section of {\sc
HAM}; {\tt\\]} introduces an external reference to some other program;
{\tt\\\\} typesets an identifier in text italics; {\tt\\|} typesets an
identifier in math italics; {\tt\\\&} typesets in boldface.

A special debugging mode is available in which \TeX\ will simply typeset
all the mini-output at the end of each section, instead of making actual
mini-indexes. This makes it easy for users to check that \CTWILL/ is
in~fact producing the information they really want. Notice that mistakes in
\CTWILL/'s output need not necessarily lead to mistakes in mini-indexes;
for example, a spurious reference in \S6 to an identifier defined
in \S5 will not
appear in a mini-index for a spread that includes \S5. It is best to make
sure that \CTWILL/'s output is correct before looking at actual
mini-indexes. Then unpleasant surprises won't occur when sections of the
program are moved from one spread to another.

When \TeX\ is finally asked to typeset the real mini-indexes, however,
it has plenty of work to do. That's when the fun begins.
\TeX's main task, after formatting the commentary and \CEE/ code
of each section, is to figure out whether the current section fits into the
current spread, and (if it does) to update the
mini-index by merging together all entries for that spread.

Consider, for example, what happens when \TeX\ typesets
\S10 of {\sc HAM}. This spread begins
with \S8, and \TeX\ has already determined that \S8 and \S9 will fit
together in a single column. After typesetting the body of \S10, \TeX\
looks at the mini-index entries. If any of them refer to \S8 or \S9, \TeX\ will
tentatively ignore them, because those sections are already part of the
current spread. (In this case that situation doesn't arise; but
when \TeX\ processed \S7, it did suppress entries for \\{vert} and \\{ark},
since they referred to \S6.) \TeX\ also tentatively discards mini-index
entries that match other entries already scheduled for the current spread.
(In this case, everything is discarded except the entries
for \\{advance} and \\{ark}; the
others---\\{next}, $t$, \\{not\_taken}, $v$, and $a$---are duplicates of
entries in the mini-output of \S8 or \S9.) Finally, \TeX\ tentatively
discards previously scheduled entries that refer to the current section.
(In this case nothing happens, because no entries from \S8 or \S9 refer to
\S10.)

After this calculation, \TeX\ knows the number~$n$ of mini-index entries
that would be needed if \S10 were to join the spread with \S8 and \S9. \TeX\
divides $n$ by the number of columns in the mini-index (here~2, but 3 in
[\TTP] and [\SGB]), multiplies by the distance between mini-baselines (here
9~points), and adds the result to the total height of the typeset text for
the current spread (here the height of $\S8+\S9+\S10$). With a few minor
refinements for spacing between sections and for the ruled line that
separates the mini-index from the rest of the text, \TeX\ is able to
estimate the total space requirement. In our example, everything fits in a
single column, so \TeX\ appends \S10 to the spread containing \S8 and \S9.
Then, after \S11 has been processed in the same fashion, \TeX\ sees that
there isn't room for \S\S8--11 all together; so it decides to begin a new
spread with \S11.

The processing just described is not built in to \TeX, of course. It is all
under the control of a set of macros called {\tt ctwimac.tex}~[\Ctwill].
The first thing \CTWILL/ tells \TeX\ is to input those macros.

\TeX\ was designed for typesetting, not for programming; so it is at best
``weird'' when considered as a programming language. But the job of
mini-indexing does turn out to be programmable.
The full details of {\tt ctwimac} are too complex to exhibit
here, but \TeX\ hackers will appreciate some of the less obvious ideas that
are used. (Non-\TeX nicians, please skip the rest of this long paragraph.) \TeX\
reads the mini-outputs of \CTWILL/ twice, with different definitions of
{\tt\\[} and {\tt\\]} each time. Suppose
we are processing section~$s$, and suppose that the current spread begins with
section~$r$. Then \TeX's
 token registers 200, 201, \dots,~219 contain all mini-index
entries from sections $r$, $r+1$, \dots,~$s-1$
for identifiers defined respectively in sections $r$, $r+1$, \dots,~$r+19$
of the \CWEB/ program. (We need not keep separate tables for
more than 20 consecutive sections starting with the base~$r$ of the current
spread, because no spread can contain more than 20 sections.) Token
register~199 contains, similarly, entries that refer to sections preceding
$r$, and token register~220 contains entries that refer to sections $r+20$
and higher.
Token register 221 contains entries for identifiers defined in other
programs. Count register $k$ contains the number of entries in token
register~$k$, for $199\le k\le221$. When count register $k$ equals~$j$, the
actual content of token register~$k$ is a sequence of $2j$ tokens,
$$\hbox{\tt\\lmda\\cs$_1$\\lmda\\cs$_2\,\ldots\,$\\lmda\\cs$_j$}$$
where each {\tt\\cs$_i$} is a control sequence defined via
{\tt\\csname...\\endcsname} that uniquely characterizes a mini-index entry.
\TeX\ can tell if a new mini-index entry agrees with another already in the
current spread by simply testing if the corresponding control sequence is
defined. The replacement text for {\tt\\cs$_i$} is the associated
mini-index entry, while the definition of {\tt\\lmda} is
$$\hbox{\tt\\def\\lmda\#1\{\#1\\global\\let\#1\\relax\}}$$
Therefore when \TeX\ ``executes'' the contents of a token register, it
typesets all the associated mini-index entries and undefines all the
associated control sequences. Alternatively, we can say
$$\hbox{\tt\\def\\lmda\#1\{\\global\\let\#1\\relax\}}$$
if we merely want to erase all entries represented in a token register. At
the end of a spread containing $p$ sections, we generate the mini-index
by executing token registers
199 and $200+p$ thru 221 using the former definition of {\tt\\lmda},
 and we also execute token registers 200 thru $200+p-1$
using the latter definition. Everything works like magic.

A bug in the original \TeX\ macros for {\tt TWILL} led to an embarrassing
error in the first (1986) printings of [\TTP] and [\MTP]: Control sequences
in token registers corresponding to sections of the current spread were not
erased; in other words, the contents of those token registers were simply
discarded, not executed with the second definition of {\tt\\lmda}.  The
effect was to make \TeX\ think that certain control sequences were still
defined, hence the macros would think that the mini-index entries were still
present; such entries were therefore omitted by mistake. Only about 3\% of
the entries were actually affected, so this error was not outrageous enough
to be noticed until after the books were printed and people started to read
them. The only bright spot in this part of the story was the fact that it
proved how effective mini-indexes are: The missing entries were sorely
missed, because their presence would have been really helpful.

The longest-fit method by which \CTWILL/'s \TeX\ macros allocate sections
to pages tends to minimize the total number of pages, but this is
not guaranteed. For example, it's
possible to imagine unusual scenarios in which sections \S100 and
\S101, say, do not fit on a single spread, while the three sections
\S100, \S101, \S102 actually do fit. This might happen if \S100 and \S101
have lots of references to variables declared in \S102. Similarly, we might
be able to fit \S100 with \S101 if \S99 had been held over from the
previous spread. But such situations
are extremely unlikely, and there is no reason to worry about them.
The one-spread-at-a-time strategy adopted by {\tt ctwimac}
is optimum, spacewise, for all practical purposes.

On the other hand, experience shows that unfortunate page breaks between
spreads do sometimes occur unless the user does a bit more fine tuning. For
example, suppose the text of \S7 in {\sc HAM} had been one line longer.
Then \S7 would not have fit with \S6, and we would have been left with a
spread containing just tiny little \S6 and lots of wasted white space. It
would look awful. And in fact, that's the reason the three statements
$$\hbox{$t\MG\\{ark}=\Lambda$; \ $v=y$; \ {\bf goto} \\{advance};}$$
now appear on a
single line of the program instead of on three separate lines: A bad break
between spreads was avoided by manually grouping those statements, using
\CWEB/'s {\tt@+} command.

One further problem needs to be addressed---the mini-indexes must
be sorted alphabetically. \TeX\ is essential for determining the
breaks between spreads (and consequently for determining the actual
contents of the mini-indexes), but \TeX\ is not a good vehicle for
sorting. The solution to this problem is to run the
output of \CTWILL/ twice through \TeX, interposing a sorting program
between the two runs. When \TeX\ processes {\tt ham.tex}, the macros of
{\tt ctwimac} tell it to look first for a file called {\tt ham.sref}. If no
such file is present, a file called {\tt ham.ref} will be written,
containing all the (unsorted) mini-index entries for each spread. \TeX\
will also typeset the pages as usual, with all mini-indexes in their
proper places but
unsorted; the user can therefore make adjustments to fix bad page
breaks, if necessary. Once the page breaks are satisfactory, a separate program
called {\sc REFSORT} is invoked; {\sc REFSORT} converts {\tt ham.ref} into
a sorted version, {\tt ham.sref}. Then when \TeX\ sees {\tt ham.sref}, it
can use the sorted data to make the glorious final copy.

For example, the file {\tt ham.ref} looks like this:
$$\tt\halign{#\hfil\cr
!1\cr
+ \\]\{GB\\\_SAVE\}4 \\\\\{restore\\\_graph\} \\\&\{Graph\}\cr
\hskip5em \$*(\\,)\$\cr
+ \\]\{GB\\\_GRAPH\}9 \\|\{u\} \\\&\{util\}\cr
\quad\rm\vdots\cr
+ \\]\{GB\\\_GRAPH\}8 \\|\{I\} \\\&\{long\}\cr
!2\cr
\quad\rm\vdots\cr
+ \\]"<stdio.h>" \\\\\{printf\} \\\&\{int\} (\\,)\cr
}$$
And the file {\tt ham.sref} looks like this:
$$\tt\halign{#\hfil\cr
\\]\{GB\\\_GRAPH\}10 \\\&\{Arc\} =\\\&\{struct\}\cr
\quad\rm\vdots\cr
\\]\{GB\\\_GRAPH\}9 \\|\{u\} \\\&\{util\}\cr
\\]\{GB\\\_GRAPH\}9 \\\&\{Vertex\} =\\\&\{struct\}\cr
\\donewithpage1\cr
\\[2 \\|\{a\} \\\&\{register\} \\\&\{Arc\} \$*\$\cr
\quad\rm\vdots\cr
\\]\{GB\\\_GRAPH\}20 \\\\\{vertices\} \\\&\{Vertex\} \$*\$\cr
\\donewithpage5\cr
}$$
Each file contains one line for each mini-index entry and one line to mark
the beginning (in {\tt ham.ref}) or end (in {\tt ham.sref}) of each spread.

\newsection{5. Conclusions}
Although \CTWILL/ is not fully automatic, it dramatically improves the
readability of large collections of programs. Therefore an author who has
spent a year writing programs for publication won't mind spending an
additional week improving the indexes. Indeed, a little extra time spent on
indexing generally leads to significant improvements in the text of any
book that is being indexed by its author, who has a chance to see
the book in a new light.

Some manual intervention is unavoidable, because a computer cannot know the
proper reference for every identifier that appears in program comments. But
experience with \CTWILL/'s change file mechanism indicates that correct
mini-indexes for large and complex programs can be obtained at the rate of
about 100 book pages per day. For example, the construction of change files
for the 460 pages of programs in [\SGB] took 5 days, during which time
\CTWILL/ was itself being debugged and refined.

Mini-indexes are wonderful additions to printed books, but we can expect
hypertext-like objects to replace books in the long run. It's easy to imagine
a system for viewing \CWEB/ programs in which you can find the meaning of
any identifier just by clicking on it. Future systems will perhaps present
``fish-eye'' views of programs, allowing easy navigation through
complicated webs of code. (See [\HYP] for some steps in that direction.)

Such future systems will, however, confront the same issues that are faced
by \CTWILL/ as it constructs mini-indexes today. An author who wants to
create useful program hypertexts for others to read will want to give hints
about the significance of identifiers whose roles are impossible or
difficult to deduce mechanically. Some of the lessons taught by \CTWILL/ will
therefore most likely be relevant to everyone who tries to design literate
programming systems that replace books as we now know them.

\newsection{References}

\tolerance=5000 \hbadness=4000
\def\ref#1. {\item{#1.\enspace}}

\ref\HYP. Brown M, Czejdo B (1990) A hypertext for literate programming.
 In: Lecture Notes in Computer
 Science, Vol.~468, 250--259

\ref\WEB. Knuth DE (1984) Literate programming.
  In: The Computer Journal, Vol.~27, No.~2, 97--111

\ref\TTP. Knuth DE (1986) Computers \& Typesetting, Vol.~B, \TeX:~The Program.
 Addison-Wesley, Reading, Massachusetts

\ref\MTP. Knuth DE (1986) Computers \& Typesetting, Vol.~D, \MF\,:~The
 Program. Addison-Wes\-ley, Reading, Massachusetts

\ref\LP. Knuth DE (1992) Literate Programming.
 CSLI Lecture Notes, No.~27, distributed by the University of Chicago Press,
 Chicago

\ref\Cweb. Knuth DE, Levy S (1990) {\tt CWEB} User Manual:
 The {\tt CWEB} System of Structured Documentation. Computer Science
 Department Report {\mc STAN}-{\mc CS}-90-1336, Stanford University, Stanford,
 California; revised version available on the Internet via anonymous ftp
 from {\tt labrea.stanford.edu} in file~{\tt \char`~ftp/pub/cweb/cwebman.tex}

\ref\SGB. Knuth DE (1993) The Stanford GraphBase: A Platform for
 Combinatorial Computing. ACM Press, New York

\ref\Ctwill. Knuth DE (1993) {\tt CTWILL}. Available
via anonymous ftp from {\tt labrea.stanford.edu} in directory
{\tt \char`~ftp/pub/ctwill}

\ref\SGBcode. Stanford University Computer Science Department (1993) The
Stanford GraphBase. Available via anonymous ftp from {\tt labrea.stanford.edu}
in directory {\tt \char`~ftp/pub/sgb}

\ref\Virg. Pharr C (1930) Virgil's {\AE}neid, Books I--VI.
 D.\,C. Heath, Boston

\ref\Ween. Weening JS (1983) Personal communication. Pre\-served in
the archives of the \TeX\ project in Stan\-ford University Library's Department
of Special Collections, SC~97, series~II, box~18, folder~7.6

\bye

\ref\HYP. M. Brown and B. Czejdo, ``A hypertext for literate programming,''
 {\sl Lecture Notes in Computer
 Science\/ \bf468} (1990), 250--259.

\ref\WEB. Donald E. Knuth, ``Literate programming,''
  {\sl The Computer Journal\/~\bf27},2 (May 1984), 97--111.

\ref\TTP. Donald E. Knuth, {\sl \TeX: The Program\/}, Volume B of
{\sl Computers \& Typesetting\/} (Reading, Massachusetts: Addison-Wesley,
 1986), {\mc ISBN} 0-20-113437-3, $\rm xvi + 594$~pp.
 Fifth printing, revised, $\rm xviii + 600$~pp., 1993.

\ref\MTP. Donald E. Knuth, {\sl{\slMF}: The Program\/}, Volume D of
 {\sl Computers \& Typesetting\/} (Reading, Massachusetts: Addison-Wesley,
 1986), {\mc ISBN} 0-20-113438-1, $\rm xvi+560$~pp.
 Third printing, revised, $\rm xviii+566$ pp., 1991.

\ref\LP. Donald E. Knuth, {\sl Literate Programming\/}
 (Stanford, California:
 Center for the Study of Language and Information, 1992),
 {\mc ISBN} 0-937073-80-6, $\rm xvi+368$~pp.
 (CSLI Lecture Notes, no.~27.) Distributed by the University of Chicago Press.

\ref\Cweb. Donald E. Knuth and Silvio Levy, {\sl {\sltt CWEB} User Manual:
 The {\sltt CWEB} System of Structured Documentation}, Computer Science
Department Report {\mc STAN}-{\mc CS}-90-1336, Stanford University, Stanford,
 CA (October 1990), 200~pp. Also published as research report GCG~23,
University of Minnesota Geometry Center, Minneapolis, MN (October 1990),
200~pp. Revised version, 226~pp., available on the Internet via anonymous ftp
from {\tt labrea.stanford.edu} in file {\tt \char`~ftp/pub/cweb/\allowbreak
cwebman.tex}.

\ref\SGB. Donald E. Knuth, {\sl The Stanford GraphBase: A Platform for
 Combinatorial Computing\/} (New York: ACM Press, 1993), {\mc ISBN}
 0-00-000000-0, $\rm viii+000$~pp.

\ref\Ctwill. Donald E. Knuth, {\sltt CTWILL}, a family of programs available
via anonymous ftp from {\tt labrea.stan\-ford.edu} in directory
{\tt \char`~ftp/pub/ctwill}.

\ref\SGBcode. Stanford University Computer Science Department, {\sl The
Stanford GraphBase}, available via anonymous ftp from {\tt labrea.stanford.edu}
in directory {\tt \char`~ftp/pub/sgb}.

\ref\Virg. Clyde Pharr, {\sl Virgil's {\AE}neid, Books I--VI\/} (Boston:
 D. C. Heath, 1930), $\rm xl+367+85$~pp.

\ref\Ween. Joseph S. Weening, personal communication, 1983. (Preserved in
the archives of the \TeX\ project in Stanford University Library's Department
of Special Collections, SC~97, series~II, box~18, folder 7.6.)

\bye